\documentclass[dvips]{article}

\usepackage{icrc2011}
\usepackage{url}

\title{Ionospheric propagation effects for UHE neutrino detection with the lunar Cherenkov technique
}

\newcommand{\etal}{\MakeLowercase{\textit{et al. }}} 
\shorttitle{McFadden \etal Ionospheric propagation effects}

\authors{Rebecca McFadden$^{1,2}$, Ron Ekers$^{2}$, Justin Bray$^{2,3}$ }
\afiliations{$^1$ASTRON Netherlands Institute for Radio Astronomy, 7991 PD Dwingeloo, The Netherlands\\ $^2$CSIRO Astronomy and Space Science, Epping, NSW 1710, Australia\\$^3$Dept. of Physics, School of Chemistry \& Physics, Univ. of Adelaide, SA 5005, Australia }
\email{mcfadden@email.astron.nl}

\abstract{Lunar Cherenkov experiments aim to detect nanosecond pulses of Cherenkov emission produced during UHE cosmic ray or neutrino interactions in the lunar regolith. Pulses from these interactions are dispersed, and therefore reduced in amplitude, during propagation through the Earth's ionosphere. Pulse dispersion must therefore be corrected to maximise the received signal to noise ratio and subsequent chances of detection. The pulse dispersion characteristic may also provide a powerful signature to determine the lunar origin of a pulse and discriminate against pulses of terrestrial radio frequency interference (RFI). This characteristic is parameterised by the instantaneous Total Electron Content (TEC) of the ionosphere and therefore an accurate knowledge of the ionospheric TEC provides an experimental advantage for the detection and identification of lunar Cherenkov pulses. We present a new method to calibrate the dispersive effect of the ionosphere on lunar Cherenkov pulses using lunar Faraday rotation measurements combined with geomagnetic field models.}

\begin{document}
\maketitle
The lunar Cherenkov technique \cite{Dagkesamanskii1989} uses the moon as a large volume detector for UHE particles interacting in the lunar regolith. This technique was pioneered by Hankins, Ekers and O'Sullivan using the 64-m Parkes radio telescope \cite{Hankins1996Asearch} and is currently being used in experiments by the NuMoon \cite{TerVeen2010,Singh2011}, RESUN \cite{Jaeger2010} and LUNASKA  \cite{James2010MNRAS} collaborations. UHE particles interacting in the Moon's regolith cause cascades of secondary particles which, due to a negative charge excess, produce coherent Cherenkov radiation \cite{Askaryan1962} that may be detected by radio telescopes. The Cherenkov radiation has a broad spectrum which peaks under a few gigahertz with the exact position of the peak determined by de-coherence and/or attenuation effects in the regolith and therefore dependent on shower geometry.

In the time domain this produces a very narrow pulse with no obvious modulation features, however, the pulse is only narrow before propagation as it becomes dispersed by the ionosphere before reaching an earth-based detector. Ionospheric dispersion destroys the coherency of the pulse which reduces the pulse amplitude and subsequent chances of detection. Pulse amplitude may be recovered using dedispersion techniques and this requires an understanding of the dispersion characteristic and current ionospheric conditions.

\section{Effects of Ionospheric Dispersion}

\begin{figure}
\begin{center}
\includegraphics [width=0.45\textwidth,keepaspectratio]{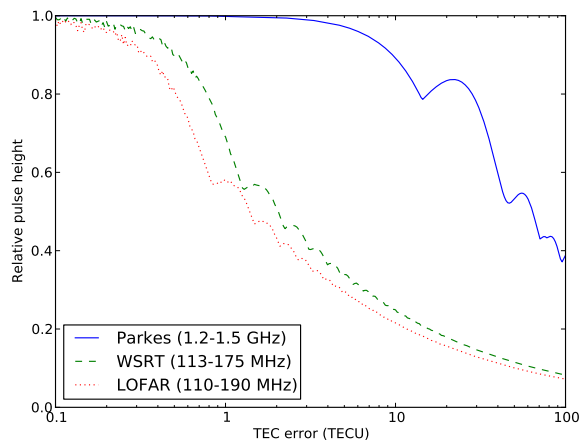}
\caption{Pulse amplitude loss due to dispersion error for the NuMoon WSRT and LOFAR experiments and the LUNASKA Parkes experiment. This assumes that dispersion was the only mechanism for loss of sensitivity and does not account for other effects such as finite sampling (see \cite{Singh2011}) and phase shifts. The Parkes curve assumes downcoversion with a high-side local oscillator at 1555 MHz, while the WSRT and LOFAR curves assume direct sampling.}\label{Disploss}
\end{center}
\end{figure}

The ionosphere is a shell of weakly ionized plasma primarily influenced by ultraviolet radiation from the sun. Its Total Electron Content (TEC) is subject to a strong diurnal cycle and is also dependent on the season, the current phase of the 11-year solar cycle and the geometric latitude of observation. The frequency-dependent refractive index of the ionospheric plasma causes a differential additive delay across the bandwidth of a propagating pulse (Equation \ref{eq:Derived3}) which results in a loss of coherency and reduction of the received pulse amplitude (Figure \ref{Disploss}).

This effect is most dramatic for low frequency experiments such as the NuMoon Westerbork Synthesis Radio Telescope (WSRT) \cite{TerVeen2010} and LOFAR \cite{Singh2011} experiments which operate in the 100-200MHz range. The differential delay is given by

\begin{equation}
\Delta t= 0.00134 \times STEC \times (\nu_{\rm{lo}}^{-2}-\nu_{\rm{hi}}^{-2}),
\label{eq:Derived3}
\end{equation}
where $\Delta t$ is the duration of the dispersed pulse in seconds, $\nu_{\rm{lo}}$ and $\nu_{\rm{hi}}$ are the receiver bandwidth band edges in Hz and $STEC$ is the Slant Total Electron Content in electrons per cm$^{2}$ which refers to the total electron content along the line of sight to a particular target object.

A reduction of the received pulse amplitude not only affects the chances of pulse detection but also the neutrino energy threshold. The amplitude of the pulse at the lunar surface is related to the charge excess in the particle shower. This excess is roughly proportional to the number of particles in the electromagnetic cascade which in turn is proportional to the energy of the shower and the energy of the original neutrino. Therefore as the minimum detectable pulse height is increased, the minimum detectable neutrino energy is also increased.

\section{Ionospheric Monitoring}

Coherent pulse dedispersion can be used to recover pulse amplitude. This can be implemented via matched filtering techniques and requires an accurate knowledge of the current ionospheric conditions, particularly the instantaneous TEC. Current methods of ionospheric monitoring include using data from Global Positioning System (GPS) satellites and ground-based ionosondes.

TEC Measurements can be derived from dual-frequency GPS signals and are available online from several sources. Until recently these values have not been made available in real time and we have made use of CDDIS TEC data \cite{CDDIS} which is sampled at two-hour intervals and published with a few days delay. More recently NASA \cite{Nasa} has made global near real-time TEC maps available at 15-minute intervals and the Australian Bureau of Meterology \cite{IPS} has done likewise for the Australasia region, although only the latter are also publicly available as data files. Both of these near real-time services produce estimates derived from GPS measurements processed with Kalman filters and combined with the IRI-2007 ionospheric model, which is driven by real-time foF2 observations from IPS ionosondes.

\begin{figure}[htb]
\begin{center}
\includegraphics [width=0.5\textwidth,keepaspectratio]{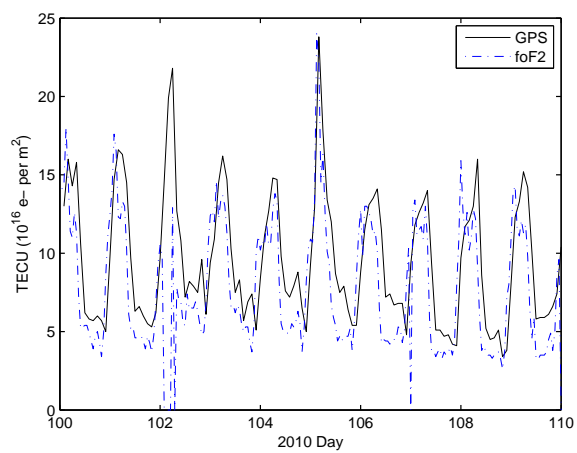}
\caption{Comparison of VTEC data from dual-frequency GPS and foF2 ionosondes.}\label{foF2_comp}
\end{center}
\end{figure}

Ionosondes probe the peak transmission frequency (fo) through the F2-layer of the ionosphere which is related to the ionospheric TEC squared. Near real-time TEC measurements can also be derived from foF2 ionosonde measurements and are available hourly from the Australian Bureau of Meteorology \cite{IPS}. However, there are known inaccuracies in the derivation of the foF2-based TEC estimates as they are empirically derived. Comparison with GPS data shows that the foF2-derived TEC data consistently underestimates GPS TEC measurements (for e.g. Figure \ref{foF2_comp}). This may be attributed to the ground-based ionosondes probing mainly lower ionospheric layers and not properly measuring TEC contribution from the plasmasphere. The error from this effect is estimated to be 1-2 TECU \cite{ho1997comparative} which has a minimal effect on high frequency experiments but for low frequency experiments this corresponds to a reduction in pulse amplitude of almost 60\%.

\begin{figure}[!tbph]
\begin{center}
\includegraphics [width=0.4\textwidth,keepaspectratio]{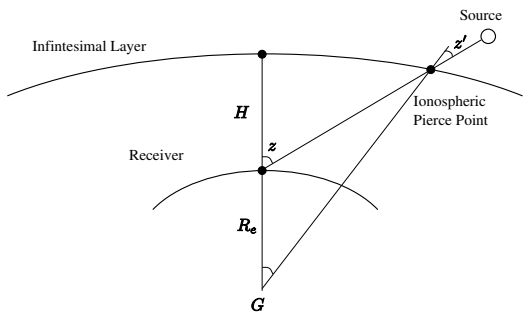}
\caption{Parameters of the ionospheric Single Layer
Model.}\label{Iono_model}
\end{center}
\end{figure}

Both of the GPS and foF2 TEC products are published as vertical TEC (VTEC) maps which must be converted to STEC estimates to obtain the true total electron content through the slant angle line-of-sight to the Moon. To perform this conversion, the ionosphere can be modeled as a Single Layer Model (SLM) \cite{Todorova2008} which assumes all free electrons are concentrated in an infinitesimally thin shell and removes the need for integration through the ionosphere. Slant and vertical TEC are related via

\begin{equation}
STEC=F(z)VTEC. \label{Eq:slant}
\end{equation}
where $F(z)$ is a slant angle factor defined as

\begin{eqnarray*}
F(z) & = & \frac{1}{\cos(z^{\prime})}\\
     & = & \frac{1}{\sqrt{1-\left(\frac{R_e}{R_e+H}\sin(z)\right)^2}}
\end{eqnarray*}
$R_e$ is the radius of the Earth, $z$ is the zenith angle to the
source and $H$ is the height of the shell (see Figure \ref{Iono_model}). The CDDIS also use
an SLM ionosphere for GPS interpolation algorithms and assume a
mean ionospheric height of 450 km. The slant conversion provides a considerable source of error, however, it is difficult to quantify as it is elevation angle dependent and influenced by horizontal gradients in the ionosphere. Combined with interpolation errors and instrumental biases this error can be as high as 3-10 TECU \cite{ho1997comparative}.

\section{A New Technique}

New methods of ionospheric monitoring are required particularly for the current low-frequency lunar Cherenkov experiments and, as the solar cycle enters a more active phase, accurate pulse dedispersion will become a more important experimental concern at all frequencies.

We have formulated a technique to obtain TEC measurements
that are both instantaneous and line-of-sight to the Moon. Ionospheric TEC can be deduced from the Faraday rotation measurements
of a polarised source combined with a geomagnetic field model,
which are more stable than ionospheric models (IGRF magnetic field values are
accurate to better than 0.01\% \cite{IGRF}). We propose to use this method with lunar thermal emission as the polarised source which is possible since Brewster angle effects produce
a nett polarisation excess in the emission from the lunar limb
\cite{Heiles1963}.

Traditional methods of planetary synthesis imaging and polarimetry require a complete set of antenna spacings and enough observing time for earth rotation synthesis. Faraday rotation measurements obtained through synthesis imaging will therefore be averaged over the entire observational period and not contain any information on short-term ionospheric structure. Depending on the chosen experimental strategy, the lunar imaging baseline requirements may also conflict with the unique constraints of a lunar Cherenkov experiment. For UHE particle detection, long antenna spacings may be preferred to minimize the level of lunar noise correlation between antennas or short spacings may be used to minimize the relative geometric delays between antennas. To overcome these limitations we have a developed a method of obtaining lunar Faraday rotation estimates in the visibility domain (i. e. without Fourier inversion to the image plane).

Working in the visibility domain relaxes the array configuration constraints and the need for earth rotation synthesis, allowing measurements to obtained in real time. This technique makes use of the angular
symmetry in the polarisation distribution of a planetary object such as the Moon. The intrinsic
thermal radiation of a planetary object appears increasingly
polarised toward the limb, when viewed from a distant point such
as on Earth \cite{Heiles1963, SPORT2002}. The polarised emission is
radially aligned and due to the changing angle of the planetary
surface toward the limb combined with Brewster angle effects. The
angular symmetry of this distribution can be exploited by an
interferometer so that an angular spatial filtering technique may
be used to obtain real-time position angle measurements directly
in the visibility domain. Measured position angles are
uniquely related to the corresponding $uv$ angle at the time of
the observation and comparison with the expected radial position
angles, given the current $uv$ angle of the observation, gives an
estimate of the Faraday rotation induced on the Moon's polarised
emission. Faraday rotation estimates can be combined with
geomagnetic field models to determine the associated ionospheric
TEC and subsequently provide a method of calibrating the current
atmospheric effects on potential Cherenkov pulses \cite{McFadden2010}.

As a preliminary verification, observations of the Moon were taken using the 22-m telescopes of
the Australia Telescope Compact Array with a center frequency of
1384 MHz. Using the angular spatial filtering technique, instantaneous position angle
estimates were calculated directly in the visibility domain of the
lunar data. Faraday rotation estimates were then obtained by comparing these to the expected $uv$ angles
and averaged over small time increments to smooth out noise-like fluctuations. Since
the received polarised lunar emission varied in
intensity over time, there were nulls during which the obtained
position angle information was not meaningful. A threshold was
applied to remove position angle measurements taken during these
periods of low polarised intensity and baseline averaging was
performed as each baseline was affected differently by the nulls. The
Faraday rotation estimates were converted to estimates of
ionospheric TEC via

\begin{equation}
\Omega = 2.36 \times 10^4 \nu^{-2}\int_{\rm{path}}N(s)B(s)\cos
(\theta) ds \label{eq:Faraday_rotation}
\end{equation} where $\Omega$ is the rotation angle in radians, $f$ is the signal
frequency in Hz, $N$ is the electron density in m$^3$, $B$ is the
geomagnetic field strength in T, $\theta$ is the angle between the
direction of propagation and the magnetic field vector and $ds$ is
a path element in m.

\begin{figure}[htb]
\begin{center}
\includegraphics [width=0.5\textwidth,keepaspectratio]{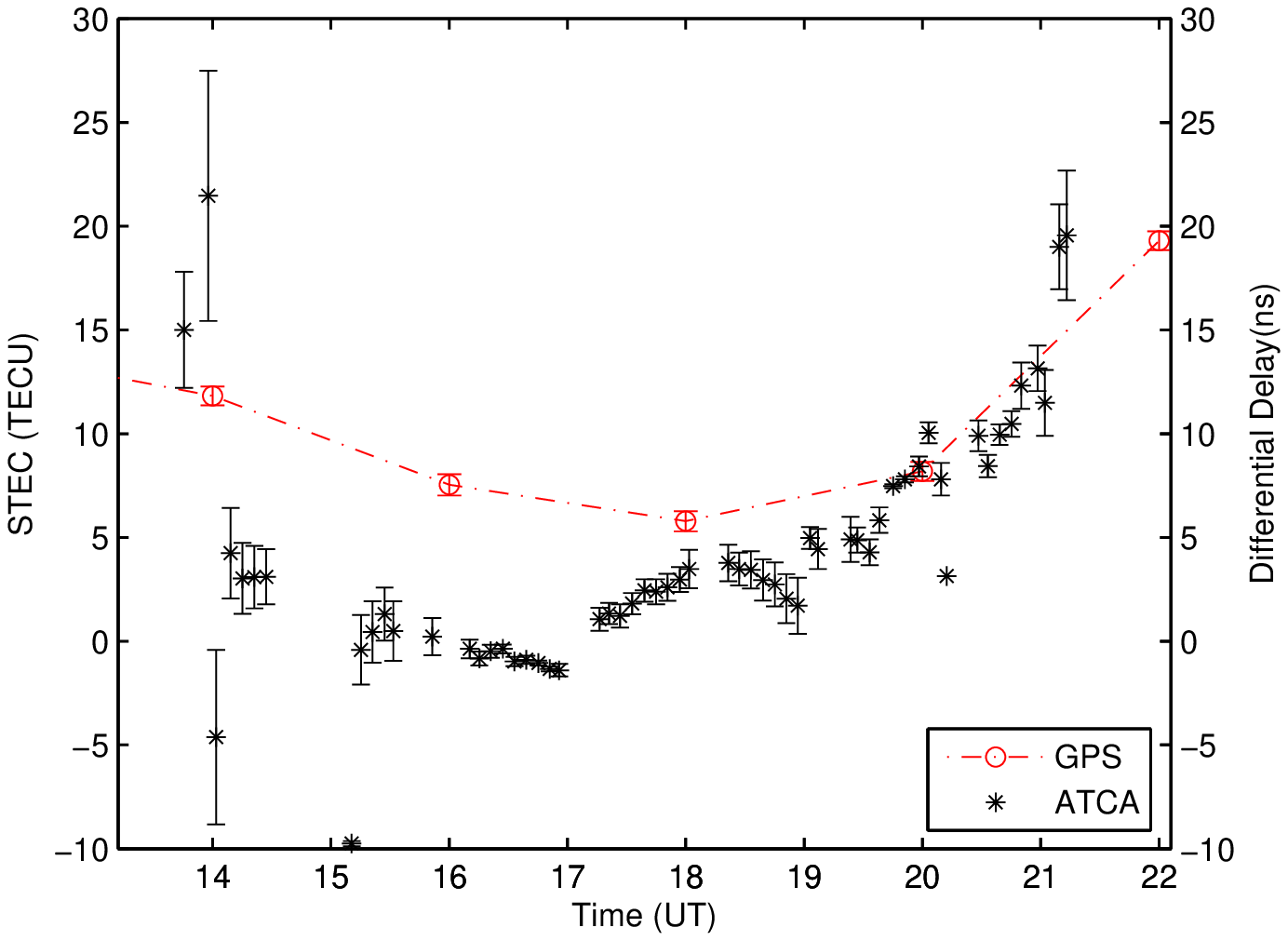}
\includegraphics [width=0.5\textwidth,keepaspectratio]{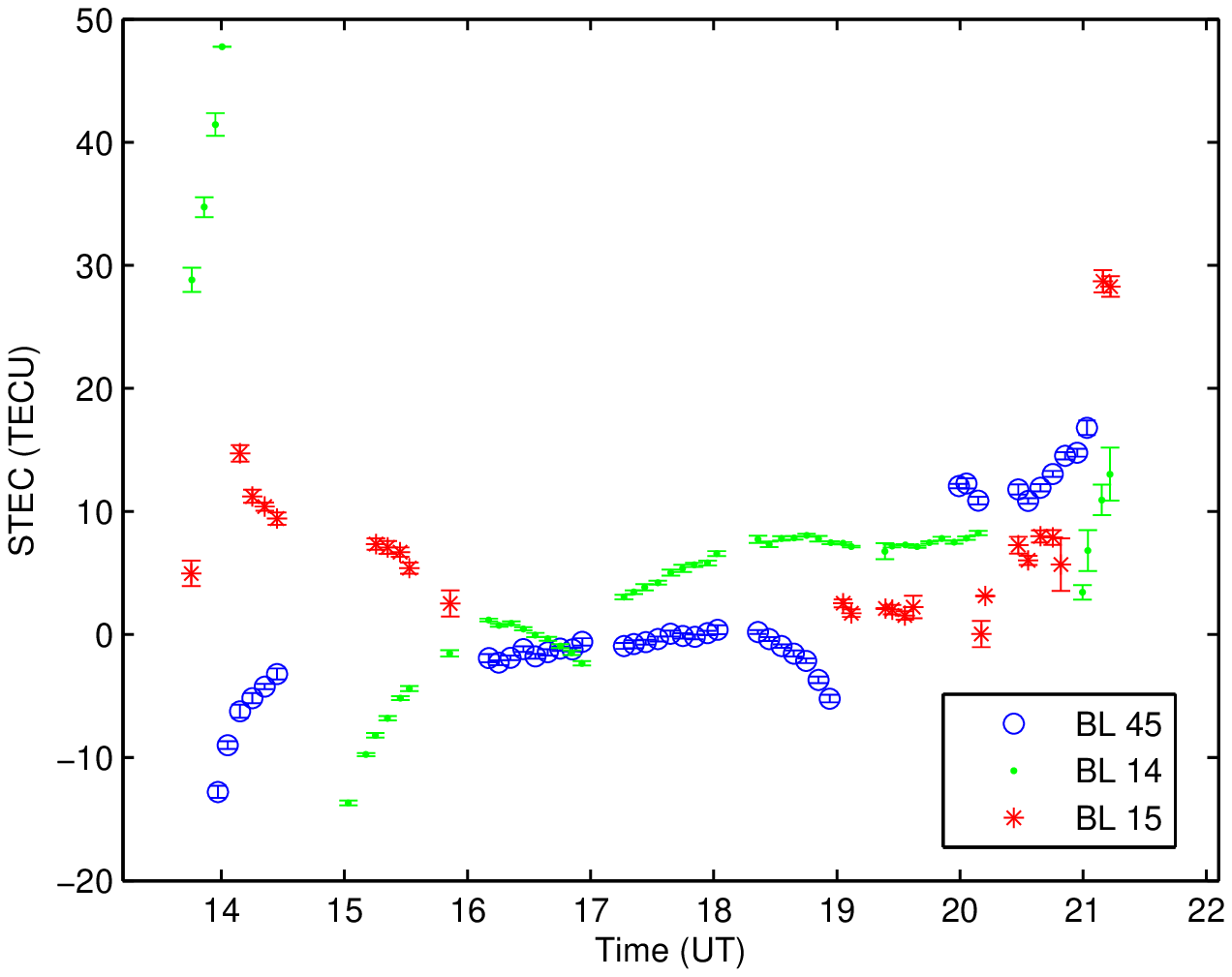}
\caption{\emph{Top}) Lunar Faraday rotation estimates converted to
(\emph{left}) ionospheric TEC values and (\emph{right}) the
differential delay across 1.2--1.8 GHz. \emph{Bottom}) TEC values before baseline averaging, for baselines 45 (61$m$), 14 (107$m$) and 15 (168$m$) }\label{TECdelay}
\end{center}
\end{figure}

To evaluate the effectiveness of this technique, the derived TEC estimates were compared against dual-frequency GPS data (Figure
\ref{TECdelay} \emph{Top}). Slant angle factors were used to convert the GPS
VTEC estimates to STEC toward the Moon. Both data sets exhibited a similar general trend,
reaching a minimum around the Moon's transit (around 16:00 UT) and rising more rapidly after transit. The steep rise in TEC after transit corresponds to a more active ionosphere at sunrise, as well as an increased path length through the ionosphere as the Moon sets. Before transit, the ionosphere was relatively stable during the night-time hours and it was mainly only the Moon's changing elevation angle which contributed to the changing TEC. The ATCA data underestimated the GPS data, particularly around 14:30--17:00 UT
where the STEC estimates may have been influenced by bad data from
the shorter baselines (Figure \ref{TECdelay} \emph{Bottom}) or due to a TEC contribution from the
plasmasphere which is not in the presence of a magnetic field
\cite{ho1997comparative}. These observations were taken when the TEC was very low and therefore the relative error in the TEC estimates is large.

\section{Conclusions}

Accurate dispersion calibration can recover pulse amplitude and increase the chances of pulse detection using the lunar Cherenkov technique. The dispersion effect is frequency dependent and strongest at low frequencies therefore low frequency experiments are the worst affected. However, as the sun enters a more active phase, accurate ionospheric pulse dispersion is becoming a more important experimental concern for lunar Cherenkov experiments at all wavelengths. Several methods of ionospheric monitoring exist including GPS and ionosonde measurements, although these methods include errors which can result in significant loss of pulse amplitude. We have presented a new technique for ionospheric calibration which uses Faraday rotation measurements of the polarised thermal radio emission from the lunar limb combined with geomagnetic field models to obtain estimates of the ionospheric TEC which are both instantaneous and line-of-sight to the Moon. Preliminary comparison to GPS data show that both data sets exhibit similar features which can be attributed to ionospheric events, however, more observations are required to investigate this technique further.


\clearpage

\end{document}